%% file: MoS2.tex
\documentclass[aps,prb,twocolumn,showpacs,amsmath,floatfix,superscriptaddress]{revtex4-1}

\usepackage{graphicx}
\usepackage{amsmath}
\usepackage{braket}
\usepackage{siunitx}

\newcommand{\MoS}[1]{MoS\ensuremath{{}_{#1}}}
\newcommand{\GWo}{}
\def\GWo/{GW$_0$}
\newcommand{\GoWo}{}
\def\GoWo/{G$_0$W$_0$}
\newcommand{\scGWo}{}
\def\scGWo/{sc-GW$_0$}
\newcommand{\angs}[1]{\ensuremath{\SI{#1}\angstrom}}

\begin{document}


\title{Screening and Many-Body Effects in Two-Dimensional Crystals: Monolayer MoS${}_2$}

\date{\today}

\author{Diana Y. Qiu}
\affiliation{Department of Physics, University of California at
Berkeley, California 94720}
\affiliation{Materials Sciences Division, Lawrence Berkeley
National Laboratory, Berkeley, California 94720}

\author{Felipe H. da Jornada}
\affiliation{Department of Physics, University of California at
Berkeley, California 94720}
\affiliation{Materials Sciences Division, Lawrence Berkeley
National Laboratory, Berkeley, California 94720}

\author{Steven G. Louie}
\email[Email: ]{sglouie@berkeley.edu}
\affiliation{Department of Physics, University of California at
Berkeley, California 94720}
\affiliation{Materials Sciences Division, Lawrence Berkeley
National Laboratory, Berkeley, California 94720}

\begin{abstract}
We present a systematic study of the variables affecting the electronic and optical properties of two-dimensional(2D) crystals within \textit{ab initio} GW and GW plus Bethe Salpeter Equation (GW-BSE) calculations. As a prototypical 2D transition metal dichalcogenide material, we focus our study on monolayer MoS${}_2$. We find that the reported variations in GW-BSE results in the
literature for monolayer MoS${}_2$ and related systems arise from different treatments of the long-range Coulomb interaction in
supercell calculations and convergence of k-grid sampling and cutoffs for various quantities such as the dielectric screening. In
particular, the quasi-2D nature of the system gives rise to fast spatial variations in the
screening environment, which are computationally challenging to resolve. We also show that
common numerical treatments to remove the divergence in the Coulomb interaction can shift the
exciton continuum leading to false convergence with respect to k-point sampling. Our findings apply to GW-BSE calculations on any low-dimensional semiconductors.
\end{abstract}


\maketitle

\section{Introduction}

Transition metal dichalcogenides (TMDs) are layered, weakly-coupled materials that can exist in few- and monolayer forms. Recently, this class of materials has attracted intense study due to the remarkable electronic and optical properties it exhibits, such as valley-selective circular dichroism, as well as coupling of spin and valley quantum numbers~\cite{cao12,xiao12,mak12} and the formation of strongly bound excitons and trions~\cite{ashwin12, qiu13, mak13, chernikov14, ugeda14, klots14, zhu14, ye14,hill15}. Molybdenum disulfide (\MoS2) is a prototypical TMD. In its most common semiconducting form (2H), monolayer \MoS2 consists of a layer of Mo atoms sandwiched between two layers of S atoms in a trigonal prismatic arrangement. In bulk and few-layer form, \MoS2 is an indirect gap semiconductor, but in monolayer form, it becomes a direct gap semiconductor, with a gap located at the $K$ and $K'$ points in the Brillouin zone~\cite{mak10,splendiani10}.

The optical spectrum of \MoS2 has been extensively studied experimentally. It has an optical gap of $1.9$~eV at room temperature~\cite{mak10, splendiani10}, which blue-shifts by as much as $0.1$~eV at low temperatures between 5 and 100K~\cite{tongay12, soklaski14}. The first peak in the optical spectrum is split by spin-orbit coupling by $0.15$~eV into two peaks commonly referred to as ``A'' and ``B''\cite{mak10}. The electronic quasiparticle bandgap is much harder to determine experimentally, but various experiments suggest that the bandgaps of monolayer \MoS2 and several other TMDs with the same structure lies between $0.2$ and $0.7$~eV above the optical gap~\cite{chernikov14, ye14, zhu14, ugeda14, klots14,zhang14,hill15}, indicating a large exciton binding energy.

There have also been numerous theoretical studies of the electronic and optical
properties of monolayer \MoS2 with widely differing results. The many-body
perturbation theory-based \textit{ab initio} GW approximation~\cite{hybertsen86} plus Bethe Salpeter equation (GW-BSE) approach~\cite{rohlfing98,albrecht98} is one the most common and accurate methods for computing quasiparticle (QP) bandstructures and optical response including electron-electron and electron-hole interactions. However, even within the general GW-BSE approach, there is a great deal of disagreement in the literature over everything from the magnitude and location of the QP bandgap to the exciton binding and excitation energies~\cite{ashwin12, lambrecht12, komsa12, wirtz13, qiu13, shi13, huser13, soklaski14,bernardi13,palummo15}. In this paper, we address the source of these inconsistencies and make note of computational issues in GW-BSE calculcations that arise for quasi-two-dimensional (quasi-2D) semiconductors and other reduced dimensional systems.

The main results of the paper are the following:

\begin{itemize}
\item The major computational challenges when dealing with mono- and
  few-layers TMDs arise from the finite extent of atomic scale in one of the
  spatial directions. This introduces rapid variations in the screening, which leads to complications in the computation of the quasiparticle and excitonic properties~\cite{qiu13,huser13}.

\item The convergence of quasiparticle gaps with respect to the k-point sampling, dielectric cutoff and number of bands included in the self-energy operator is much slower than what is reported in earlier work and is closely tied to the supercell size used and the treatment of the quasi-2D behaviour of the Coulomb interaction. The lack of convergence is sufficient to explain the varying results in the literature for GW-BSE calculations on monolayer \MoS2 and other TMDs.

\item We show that different numerical treatments of the divergence in the Coulomb interaction shifts the exciton continuum and can lead to false convergence of the binding energy with respect to k-point sampling. In particular, we find that it is possible to obtain an apparent agreement of the calculated optical gap with experiment, even though the exciton binding energy and the higher exctionic states are not computed correctly.
\end{itemize}


This paper is organized as follows. In section II, we discuss the dielectric screening in quasi-2D semiconductors and review the effect of the truncation of the Coulomb potential. In section III, we discuss the QP bandstructure, the convergence of the self-energy, including special considerations for quasi-2D systems, the effect of updating the Green's function $G$ in the $GW_0$ approach and the frequency dependence of the screening. In section IV, we discuss the effect of screening on the optical response of \MoS2, characterize the excitons and their wavefunctions, discuss how they converge in our calculations, and discuss the effects of quasiparticle lifetimes. We conclude in section V by summarizing our results.

%
%
\section{Electron-electron and electron-hole Interactions and Screening in 2D}

%
%
\subsection{\label{sect:truncation}Coulomb Truncation and Convergence}

First-principles calculations using planewave basis sets require periodic boundary conditions. This means that for 2D systems, such as monolayer \MoS2, it is necessary to increase the dimension $L_z$ of the unit cell in the aperiodic direction to avoid interactions between repeated monolayers~\cite{schluter75}. With conventional DFT functionals, such as LDA or GGA, there are no long range interactions for a neutral system, so a vacuum of $\sim\angs{5}$ ($L_z \sim \angs{10}$) is sufficient to converge the relative eigenvalues (other values, such as the work function and ionization energies, require a larger vacuum to prevent interactions between periodic images). However, when we compute the polarizability and related quantities in the GW approach, we end up calculating a response function that is long ranged, and it becomes computationally unfeasible to include enough vacuum to prevent periodic images from interacting.

One effective solution for this problem is to explitly truncate the Coulomb
interaction in real space along the aperiodic direction. This is implemented in the BerkeleyGW package~\cite{jdeslip11BGW} following Ismail-Beigi's scheme~\cite{beigi06}. The truncated Coulomb potential has a closed form in reciprocal space,
\begin{equation}
\label{eq:vT}
v^{\mathrm{trunc}}(\mathbf{q}) = \frac{4\pi}{q^2}\left[1-e^{-\frac{q_{xy} L_z}{2}}\cos\left(\frac{q_z L_z}{2}\right)\right],
\end{equation}
where $q_{xy} = (q_x^2+q_y^2)^{1/2}$. This allows us to directly compute the static RPA inverse dielectric matrix without spurious interactions between the repeated monolayers in our supercell geometry as
\begin{equation}
\epsilon^{-1}_{\mathbf{GG'}}(\mathbf{q}) =  \delta_\mathbf{GG'} + v^{\mathrm{trunc}}(\mathbf{q+G})\chi_\mathbf{GG'}(\mathbf{q}),
\end{equation}
where $\chi_\mathbf{GG'}(\mathbf{q})$ is the static non-interacting RPA polarizability.

\begin{figure}
    \includegraphics[width=246.0pt]{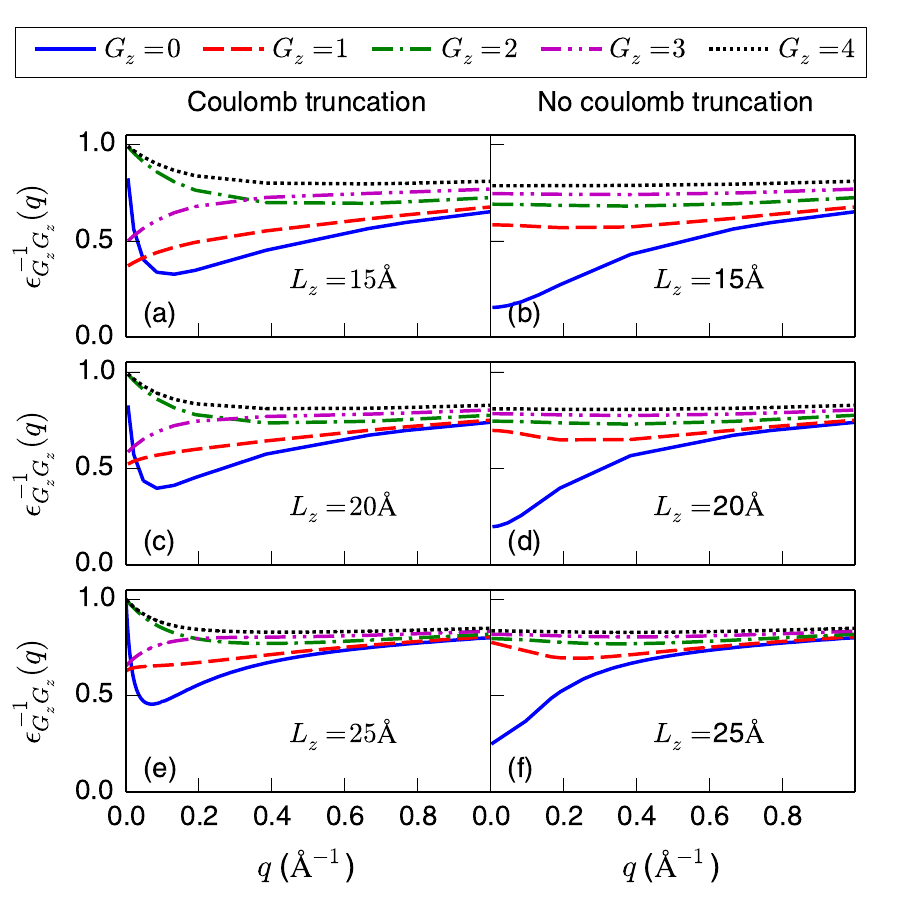}
    \caption{\label{fig:epsinv} (Color online) Evolution of the first few diagonal elements of the inverse dielectric matrix, $\epsilon^{-1}(\mathbf{q})$, for $G_x{=}G_y{=}G_x'{=}G_y'{=}0$ and $G_z{=}G_z'$ with (left) and without (right) Coulomb truncation for $L_z{=}\angs{15}$ (a,b), \angs{20} (c,d), and \angs{25} (e,f) supercell sizes. A cutoff of $35$~Ry and $6,000$ bands was used for calculating all $\epsilon^{-1}_{\mathbf{G,G'}}(\mathbf{q})$ in this figure. The value of $G_z$ is given in units of $\frac{2\pi}{L_z}$.
    }
\end{figure}

We now examine how the features of the dielectric matrix evolve with supercell size with and without Coulomb truncation. In isotropic bulk systems, the screening is dominated by the ``head'' element $\mathbf{G}{{=}}\mathbf{G'}{{=}}0$~\cite{hybertsen85,hybertsen86,hybertsen87,zhang89}. In quasi-2D systems, however, the $G_z$'s (the reciprocal lattice vectors along the aperiodic direction) are almost continuous, so it is no longer reasonable to look at the single element $G_z{=}0$. In Fig.~\ref{fig:epsinv}, we plot $\epsilon^{-1}_{\mathbf{GG}'}(\mathbf{q})$ for elements where $G_x{=}G_x'{=}G_y{=}G_y'{=}0$ and $G_z{=}G_z'$ for several different values of $G_z$. When the truncated Coulomb interaction is used, the behavior of $\epsilon^{-1}_{\mathbf{GG}'}(\mathbf{q})$ changes depending on whether $G_z$ is odd or even. $\epsilon^{-1}(\mathbf{q})$ goes smoothly to a value less than $1$ as $q$ goes to $0$, when $G_z$ is odd, and sharply returns to $1$ as $q$ goes to $0$, when $G_z$ is even. This behavior arises from the $\cos$ term in the truncated Coulomb interaction, and contrasts with the untruncated case, where $\epsilon^{-1}(\mathbf{q})$ goes to a number less than $1$ as $q$ goes to $0$ for all $G_z$'s, with most of the screening coming from $G_z{=}0$.

The screening behavior with and without Coulomb truncation also depends, unsurprisingly, on the amount of vacuum $L_z$. In both cases, consecutive $G_z$'s for $G_z>0$ become more similar as $L_z$ increases, since the separation between $G_z$'s is $\frac{2\pi}{L_z}$. Consequently, the number of $G_z$'s required to capture the screening behavior increases proportionally with $L_z$.

\begin{figure}
    \includegraphics[width=246.0pt]{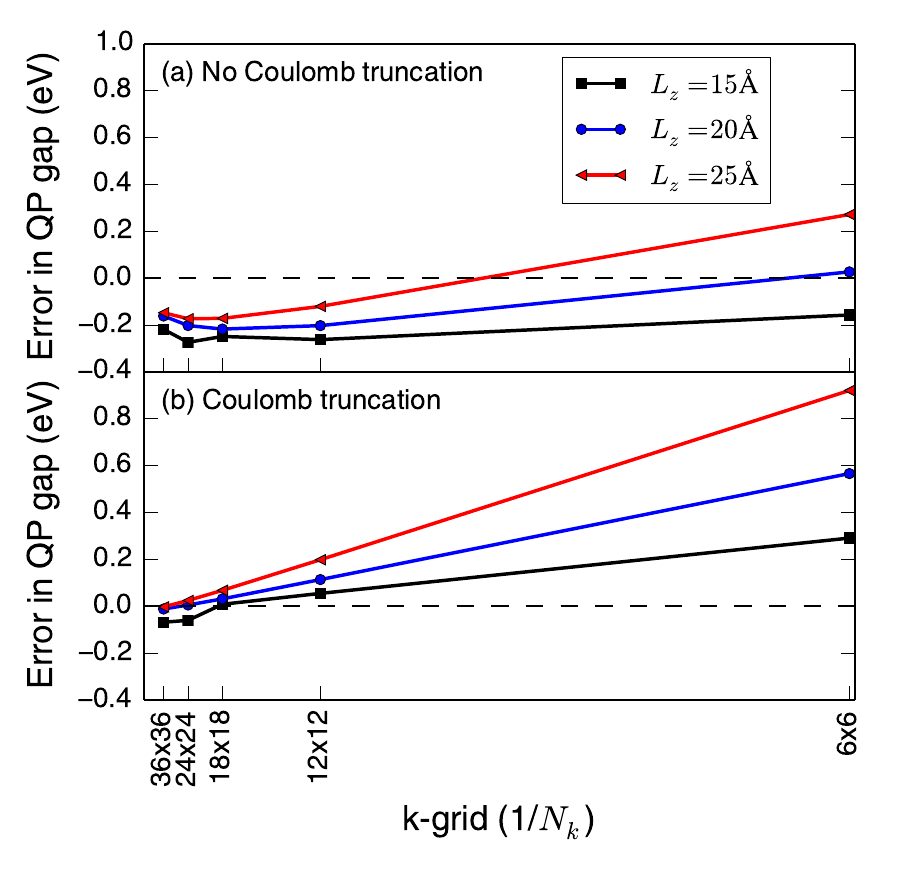}
    \caption{\label{fig:kconv} (Color online) Convergence of the error in the QP gap with k-point sampling (a) without Coulomb truncation and (b) with Coulomb truncation, for supercell sizes $L_z=\angs{15}$ (black squares), $L_z=\angs{20}$ (blue circles), and $L_z=\angs{25}$ (red triangles). Zero is set to the QP gap with Coulomb truncation extrapolated to infinite k-point sampling for $L_z=\infty$.}
\end{figure}

There is also a direct correlation between the $q$-dependence of the dielectric matrix with $L_z$ when we employ the truncated Coulomb interaction. As shown in the left panels in Fig.~\ref{fig:epsinv}, the ``dip'' feature for even $G_z$'s becomes sharper as $L_z$ increases, so the k-point sampling must be fine enough to resolve the features in $\epsilon^{-1}_{00}(\mathbf{q})$. An important consequence is that \emph{convergence of k-point sampling is tied to the size of the supercell}. Fig.~\ref{fig:kconv} shows the convergence of the QP gap with respect to k-point sampling and the size of the vacuum. When Coulomb truncation is used (Fig.~\ref{fig:kconv}~(b)), the QP gap converges more slowly for larger $L_z$'s, reflecting the need to resolve sharper features in $\epsilon^{-1}_{00}(\mathbf{q})$. However, the QP gap converges to the same value regardless of the supercell size when Coulomb truncation is used.

The picture is different and shows a significantly slower k-point convergence when we don't employ Coulomb truncation. As
shown in Fig.~\ref{fig:kconv}~(a), the QP gap still displays a very strong
dependence on k-point sampling at the densest grid size of $36\times36$. There are two important differences here with respect to
the case with truncated Coulomb potential: (1) these calculations converge to a
smaller incorrect QP gap, and (2) the convergence with respect to k-point sampling is not
monotonic but changes direction as the k-grid sampling becomes finer. Both
these facts are understood from the long wavelength behavior of the
screening. Whenever $q \lesssim 1/L_z$, the calculation without a truncated
Coulomb potential includes a spurious polarization due to the repeated
monolayers in the aperiodic direction. This spurious term screens out the
Coulomb interaction and decreases the QP gap.


Finaly, it's important to mention the dependence of the number of bands needed for the various quantities in the GW calculation on
$L_z$. The number of empty states included in our calculation is well
approximated by the number of planeawaves $\ket{\mathbf{G}}$ with kinetic energy
less than the dielectric cutoff $E=|\mathbf{G}|^2/2$, so it is proportional to
the supercell volume. If the number of bands is kept constant while $L_z$ is
increased, the screening will not be captured properly in the GW calculation,
and the QP gaps will be overestimated. We attritube the reason why some studies
found that the QP gaps increase much more when the vacuum is increased to this
false convergence~\cite{huser13,komsa12}.

%
%
\subsection{Effective 2D Dielectric Function}

For simplicity, we discuss here the static dielectric function. The same discussion caries over for the dynamic case. In general, the dielectric function of a material is defined as the following relation between the bare Coulomb potential v and the effective screened Coulomb interaction W:
\begin{equation}
\label{eq:epsinv}
W(\mathbf{r}_1, \mathbf{r}_2) \equiv \int \mathrm{d}^3 r_3\;
\epsilon^{-1}(\mathbf{r}_1, \mathbf{r}_3) v(|\mathbf{r}_3-\mathbf{r}_2|).
\end{equation}

Our goal now is to define an effective 2D dielectric function between two electrons in a monolayer material. Due to confinement, the modulus squared of the wavefunction (in a tight-binding framework) associated to either electron, $l=1,2$, can be written as $\rho_i(\mathbf{r}-\mathbf{s}_l)$, where $i=1,2$ labels different orbitals and $\mathbf{s}_l$ is a coordinate in the xy-plane around which the orbital is centered. In analogy to Eq.~\ref{eq:epsinv}, we define the effective 2D inverse dielectric function in terms of the strength of the electronic interaction integration between orbitals $i$ and $j$ as
\begin{align}
\label{eq:epsinv2D_1}
W_{i j}(\mathbf{s}_1, \mathbf{s}_2) 
&\equiv \int \mathrm{d}^3 r_1 \; \mathrm{d}^3 r_2 \;
\rho_i(\mathbf{r}_1-\mathbf{s}_1) W(\mathbf{r}_1, \mathbf{r}_2) \rho_j(\mathbf{r}_2-\mathbf{s}_2) \notag \\
&\equiv \int \mathrm{d}^2 s_3 \;
\left(\epsilon^{-1}_\mathrm{2D}\right)_{i j}(\mathbf{s}_1, \mathbf{s}_3) v(|\mathbf{s}_3-\mathbf{s}_2|).
\end{align}

In order to gain further insight on the form of the response function, we assume that both $W$ and $\epsilon^{-1}_\mathrm{2D}$ are isotropic and depend only on $s\equiv|\mathbf{s}_2-\mathbf{s}_1|$. Such a simplification allows us to write the strength of the electronic interaction between the two orbitals in real space as
\begin{equation}
\label{eq:Ws}
W_{i j}(s) = \frac{1}{2\pi L_z} \mathcal{F}_0
\left[ \sum_{\mathbf{G}_z \mathbf{G}_z'}
\rho_i^*(\mathbf{q}+\mathbf{G}_z) W_{\mathbf{G}_z \mathbf{G}_z'}(q) \rho_j(\mathbf{q}+\mathbf{G}_z')
\right]\!(s),
\end{equation}
where $\rho(\mathbf{q}+\mathbf{G}_z)\equiv\int \mathrm{d}^3r e^{i(\mathbf{q}+\mathbf{G}_z)\cdot\mathbf{r}}\rho(\mathbf{r})$, and $\mathcal{F}_0 [f](s) \equiv 2\pi\int_0^\infty \mathrm{d}q\; q\,f(q)\,J_0(q s)$ is the Hankel transform of $f$.

In reciprocal space, the effective 2D inverse dielectric function is simply the
ratio between the 2D screened Coulomb interaction (2D Fourier transform of
Eq.~\ref{eq:Ws}) and the truly two-dimensional bare Coulomb potential, $v_\mathrm{2D}(q) = 2\pi e^2/q$.
The simplest choice of orbitals is a delta function at $\mathbf{r}=(\mathbf{s},z=0)$, which yields the effective 2D screening\begin{equation}
\label{eq:epsinv2D_2}
\epsilon^{-1}_\mathrm{2D}(q)
 = \frac{q} {2\pi e^2 L_z} \sum_{\mathbf{G}_z \mathbf{G}_z'} W_{\mathbf{G}_z \mathbf{G}_z'}(q).
\end{equation}

Eq.~\ref{eq:epsinv2D_2} defines an effective 2D dielectric for a quasi-2D material,
where the complicated details of the screening in the out-of-plane direction $z$
have been integrated out. We note that our expression for $\epsilon^{-1}_\mathrm{2D}(q)$
differs from that defined in Refs.~\cite{huser13,latini15}, who define
it by the field in a region in the slab induced by a plane-wave external potential.
In contrast, Eq.~\ref{eq:epsinv2D_2} measures how
much the bare 2D Coulomb potential $v_\mathrm{2D}(q) = 2\pi e^2/q$ between two
point charges in the middle of the \MoS2
plane gets screened due to electronic screening. This is the relevant quantity to derive
low-energy Hamiltonians to model electron-electron and electron-hole interactions in
quasi-2D systems, including excitonic states and electron scattering.

\begin{figure}
    \includegraphics[width=246.0pt]{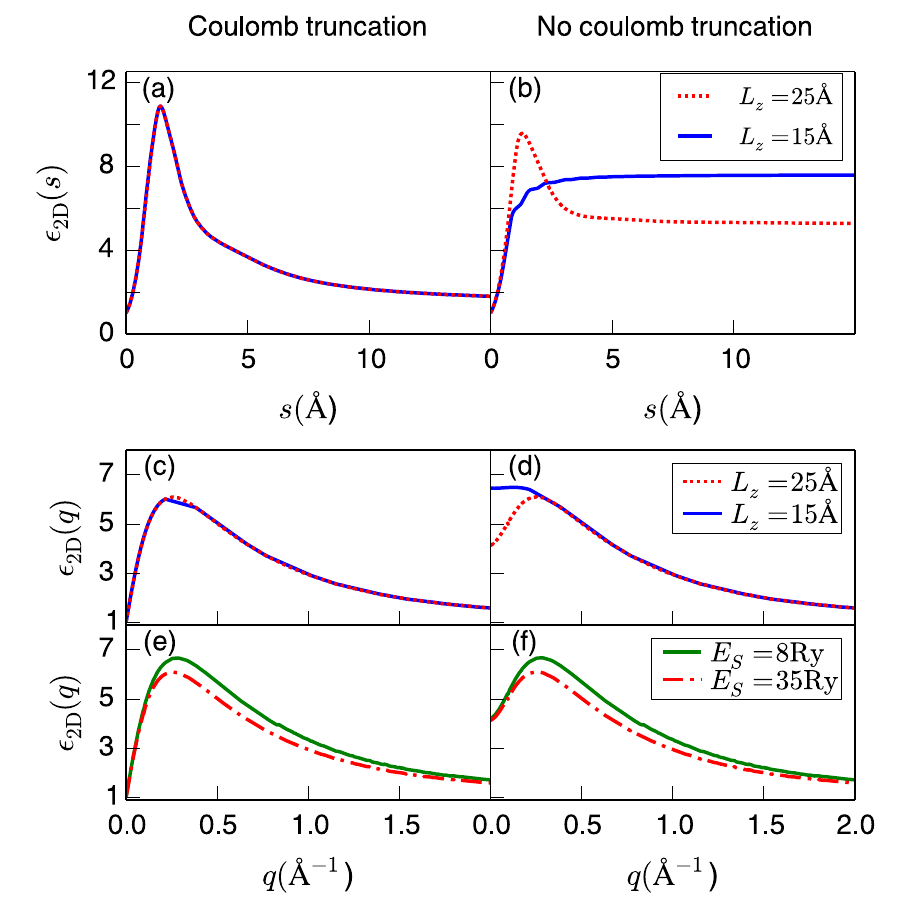}
    \caption{\label{fig:eps2D} (Color online) Effective 2D screening between two point charges in the Mo plane with Coulomb truncation (left) and without Coulomb truncation (right). Panels (a) and (b) compare the effective screening in real space when $L_z=\angs{15}$ (solid blue) and $L_z=\angs{25}$ (dotted red) with a $35$~Ry cutoff. Panels (c) and (d) are the corresponding reciprocal space plots of (a) and (b). Panels (e) and (f) compare effective screening in reciprocal space when the cutoff is $35$~Ry (dashed red) and $8$~Ry (solid green) with $L_z=\angs{25}$.}
\end{figure}

In Fig.~\ref{fig:eps2D}~(c-f), we show the reciprocal-space effective 2D
dielectric function $\epsilon_\mathrm{2D}(q)$. The corresponding real-space
curves are obtained by taking the Hankel transform of Eq.~\ref{eq:epsinv2D_2}
and are shown in Fig.~\ref{fig:eps2D}~(a,b). There is a very sharp peak in
$\epsilon_\mathrm{2D}(s)$ at $s=\angs{1.5}$, which corresponds to roughly
half the thickness of the slab. This peak can be understood if we consider
the Coulomb interaction between two point charges embeded in a quasi-2D
semiconductor: as in 2D semiconductors, if two charges are very close together,
there is not enough space for the electronic cloud to polarize, so
$\epsilon_{\mathrm{2D}}(s{\rightarrow}0) = 1$. At the same time, if the two charges
are very far away, the field lines connecting the charges travel mainly through the vacuum,
so they are not much affected by the intrinsic dielectric environment of the quasi-2D semiconductor
and $\epsilon_{\mathrm{2D}}(s{\rightarrow}\infty) = 1$. Therefore, there is a finite
distance $s_\mathrm{max}$ where $\epsilon_{\mathrm{2D}}(s_\mathrm{max})$ must exhibit
its maximum. The value of the peak of $\epsilon_{\mathrm{2D}}(s_\mathrm{max})$ depends on
the polarizability and thickness of the material. We note that $L_z$ should have no effect on the effective 2D screening as long as it
is large enough to contain the charge density within the truncated Coulomb interaction approach. This is \textit{not} true for the untruncated case.


For very short distances ($s < \angs{1}$), the effective 2D dielectric screening
with and without truncation are similar, but at larger distances polarizability of the replica slab together with the long-range interaction results in drastic overscreening. Instead of
approaching $1$, $\epsilon_{\mathrm{2D}}(s)$ approaches a larger finite
constant, which is the macroscopic dielectric constant of a bulk system consisting of layers of \MoS2 separated by layers of vacuum. While this constant indeed approaches $1$ as $L_z \rightarrow \infty$, it does so very slowly. Thus, it is very important to truncate the Coulomb interaction to include correctly the effects of the dielectric response of quasi-2D systems.

Similar features are seen in the effective 2D dielectric function for the converged results in reciprocal space, as shown in Fig.~\ref{fig:eps2D}~(c,d). Specifically: (1) there is a peak in $\epsilon_{\mathrm{2D}}(q)$; (2) when the Coulomb interaction is truncated, $\epsilon_{\mathrm{2D}}(q)$ does not depend on $L_z$; and (3) while $\epsilon_{\mathrm{2D}}(q{\rightarrow}0)=1$ when we truncate the Coulomb potential, it incorrectly approaches a different and larger value when we don't truncate the potential.

We also show the effective screening for different energy cutoffs for
the dielectric matrix, in Fig.~\ref{fig:eps2D}~(e,f). The effect of changing the
dielectric cutoff is similar for both the truncated and untruncated Coulomb
interactions. For very small $q$'s, before the peak, screening does not depend
strongly on the cutoff. For larger $q$'s, decreasing the cutoff results in
overscreening. Therefore, depending on the property one is interested in
(quasiparticle or excitonic levels), different convergence parameters may have to be
used. In particular, the convergence of quasiparticle states, as computed within
the GW approximation, converges very slowly because the self energy depends on
$\epsilon$ at both short and long distances. 

\begin{figure}
    \includegraphics[width=246.0pt]{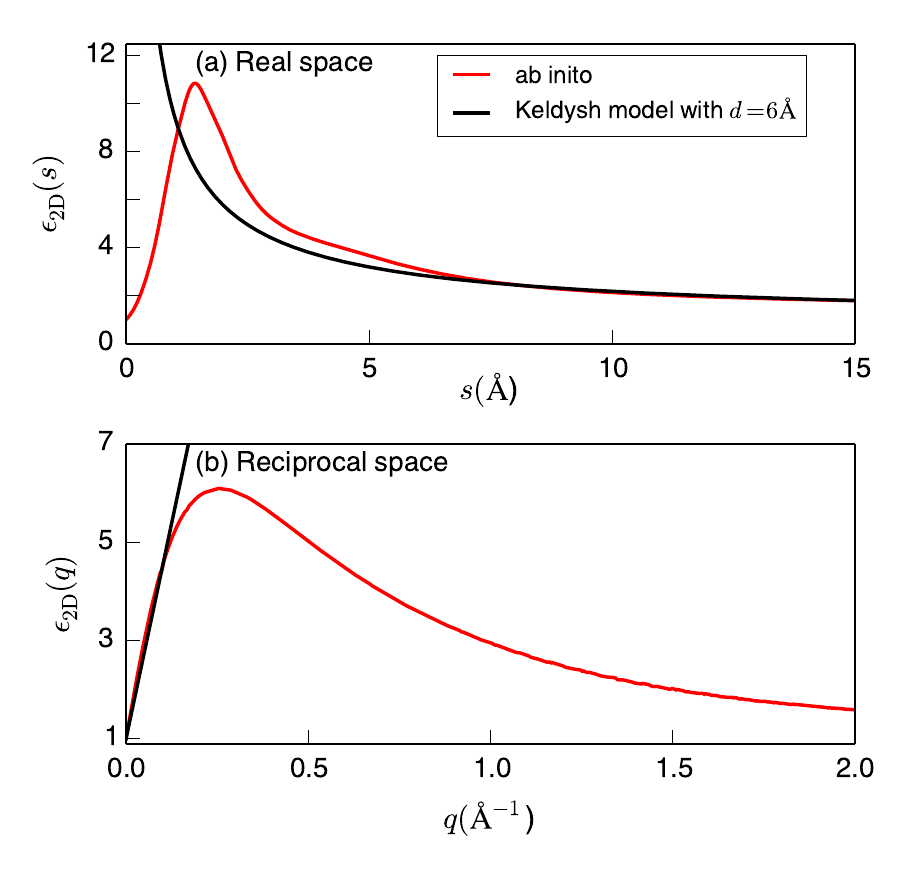}
    \caption{\label{fig:keldysh} (Color online) Comparison of the effective 2D screening as defined by Eq.~\ref{eq:epsinv2D_2} (red lines) with the Keldysh model (black lines) in real space (a) and reciprocal space (b). The Keldysh model uses an effective slab thickness of $d=6\mathrm{\AA}$ to obtain the best fit to the \textit{ab initio} results.}
\end{figure}

Finally, we compare the effective 2D screening obtained from our \textit{ab initio} calculations with the screening model developed by Keldysh~\cite{keldysh}, which is frequently used to describe screening of excitons in quasi-2D materials~\cite{cudazzo11,pulci12,berkelbach13,chernikov14,wu15,latini15}. In the Keldysh model, which is based on a slab of constant dielectric value, the potential between two charges in a slab of thickness $d$ has the form
\begin{equation}
\label{eq:keldysh}
V_{\mathrm{2D}}(s)=\frac{\pi e^2}{2\rho_0}\left[H_0\left(\frac{s}{\rho_0}\right)-Y_0\left(\frac{s}{\rho_0}\right)\right],
\end{equation}
where $H_0$ and $Y_0$ are respectively the Struve and Bessel functions of the second kind and $\rho_0$ is a screening length, which is $\rho_0=\frac{d\epsilon}{2}$~\cite{keldysh}, where $\epsilon$ is the in-plane dielectric constant of the bulk material. If the slab is taken to be strictly 2D, it has been shown~\cite{cudazzo11} that the screening length is proportional to the 2D polarizability of the layer, and taking the 2D Fourier transform of Eq.~\ref{eq:keldysh} results in a dielectric function of the form
\begin{equation}	
\epsilon_{\mathrm{2D}}(q) = 1 + \rho_0 q,
\end{equation}
where $\rho_0=2\pi\alpha_{\mathrm{2D}}$. Here, $\alpha_{\mathrm{2D}}$ is the 2D polarizability and can be related to the polarizability of the actual quasi-2D slab by fitting to the long wavelength limit of the \textit{ab initio} polarizability. We fit the Keldysh model to our \textit{ab intio} effective dielectric function at small $q$, as defined in Eq.~\ref{eq:epsinv2D_2}, and obtain an effective screening length of $\rho_0=35\mathrm{\AA}$ or an effective slab thickness of $d=6\mathrm{\AA}$, which is about twice the thickness of monolayer \MoS2 measured from the center of the sulfur atoms. A comparison of our \textit{ab intio} effective dielectric function with the best fit to the Keldysh model is shown in Fig.~\ref{fig:keldysh}. We see that the Keldysh model can be adjusted to give a good description of the form of the screening in the long wavelength limit and thus can describe the screening seen by excitons as long as the exciton radius is on the order of or larger than the screening length $\rho_0$, which is unknown without an \textit{ab initio} calculation. Moreover, for phenomena that depend on short-range or varying length scale screening, the Keldysh model would drastically overestimate the screening in quasi-2D systems.

%
%
\section{Quasiparticle Bandstructure}

In this section, we discuss the computational details and results of our GW calculation of the QP bandstructure.

\subsection{Computational Details and Convergence}

We use density functional theory (DFT)~\cite{hohenberg64,kohn65}, as implemented in Quantum ESPRESSO~\cite{espresso},
 in the local density approximation (LDA) to obtain a mean-field starting point for our
 GW calculation~\cite{hybertsen86}. Different choices of the DFT functional and a relaxed versus
 experimental crystal structure can result in about $0.1$~eV difference in the QP gap of \MoS2{}.
 We find that relaxing the structure with an LDA functional increases the gap at K
 by $0.04$~eV compared to the experimental structure. Given identical structures,
 using a GGA functional decreases the gap by $0.03$~eV compared to LDA.

We use norm-conserving pseudopotentials and include the Mo $4s$ and $4p$ semicore states
 and the $4d$ valence state. Including the semicore $4s$ and $4p$ states is necessary
 to accurately capture the exchange contribution to the self energy. However, these deep
 $4s$ and $4p$ states are not included in the charge density used in the Hybertsen-Louie
 Generalized Plasmon Pole (HL-GPP) model~\cite{hybertsen86} to calculate the self energy,
 since they are more than $35$~eV below the Fermi energy and, thus, do not
 contribute to low-energy screening. We use a supercell with \angs{25} of vacuum in the aperiodic
 direction, and we relax the supercell using a wavefunction cutoff of $350$~Ry and a
 24x24x1 k-grid, resulting in an in-plane lattice constant of \angs{3.15}, which deviates less
 than $1$\% from the experimental lattice constant of few-layer \MoS2~\cite{young68}.
 Then, we generate wavefunctions used in the GW-BSE calculation using a wavefunction\cite{hybertsen86}
 cutoff of $125$~Ry, which is sufficient to converge the bare exchange contribution~
 to the QP gap to within $0.01$~eV.

Our GW calculation is performed with the BerkeleyGW package~\cite{jdeslip11BGW}.
 We calculate the dielectric matrix using the truncated Coulomb interaction discussed
 in section II and using a  24x24x1 k-point sampling to converge the QP gap to within
 $0.05$~eV (see Fig.~\ref{fig:kconv}). We take into account dynamical screening effects
 in the self energy through the HL-GPP model. We also use the 
 static remainder technique~\cite{jdeslip10SR} to reduce the number of necessary unoccupied states.

\begin{figure}
    \includegraphics[width=246.0pt]{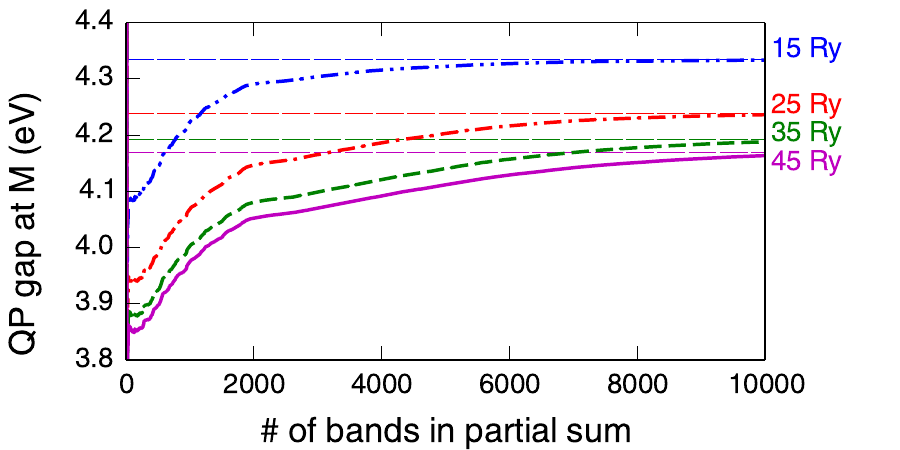}
    \caption{\label{fig:QPconverge} (Color online) Convergence of the QP gap at the M point with respect to the number of bands included in the partial sum for the Coulomb-hole contribution to the self energy, for dielectric cutoffs of $15$(blue), $25$(red), $35$(green), and $45$(magenta)~Ry. The static remainder correction is included. The dashed lines indicate the value of the QP gap extrapolated to infinite bands.}
\end{figure}

As discussed in our previous work~\cite{qiu13}, GW calculations on \MoS2 and TMDs in
 general converge very slowly with respect to the energy cutoff ($E_S$) of the dielectric
 matrix and the number of bands ($N_b$) included in the polarizability and Coulomb-hole
 summations of the self energy. The slow convergence of $E_S$ arises from the presence of localized $d$ orbitals
 near the Fermi energy and the different character of the valence and conduction bands.
 The slow convergence of $N_b$ arises due to the large number of G-vectors in the dielectric
 matrix and the supercell size, as discussed in section II.A. Our calculation required $N_b=6000$
 bands and a dielectric cutoff of $E_S=35$~Ry to converge the QP gaps to better than $0.05$~eV,
 for a total error bar of $\sim 0.1$~eV when combined with the error bar due to k-point sampling.
 To test the convergence of the number of bands we calculated QP gaps with a dielectric cutoff
 of up to $E_S=45$~Ry and up to $N_b=12000$ bands (Fig.~\ref{fig:QPconverge}).

As Shih \textit{et al.}~\cite{shih10} have noted, the dielectric cutoff and bands are interdependent parameters and
 attempting to converge the number of bands using a dielectric cutoff that is too small or converge the
 dielectric cutoff using too few bands will result in false convergence. The static remainder technique
 speeds up convergence considerably when only a few bands are included, but for a precision
 of greater than $0.1$~eV, the convergence with respect to bands for a fixed $E_S$ is about the
 same with and without static remainder. The static remainder is still helpful, however,
 because when using static remainder, convergence with respect to bands is in the opposite
 direction as convergence with respect to $E_S$, resulting in some cancellation of error.

We also self-consistently update the eigenvalues of the Green's function, $G$, when building
 the self-energy operator $\Sigma$. We find that going to $G_1W_0$ increases the QP gap at
 $K$ by $0.08$~eV compared to $G_0W_0$. Further updating $G$ increases the QP gap at $K$ by
 only $0.02$~eV, so we stop at the $G_1W_0$ level. The bandgap is 2.59 eV at the $G_0W_0$ level and
 2.67 eV at the $G_1W_0$ level, with spin-orbit interactions included.
 
We also compare results obtained using the HL-GPP model with the full-frequency dielectric matrix calculated using the contour-deformation approach~\cite{contourdef,contourdef2}. At the $G_0W_0$ level, the full-frequency bandgap is 2.45 eV and increases to 2.54 eV after self-consistently updating the eigenvalues in $G$. Thus, inclusion of the explicit dynamical effects decreases the gap by 0.13 eV compared with the HL-GPP.
 
We include spin-orbit as a perturbation, and find that the valence band at $K$ is split by $0.15$~eV. The details of the implementation are discussed in section IV.A.3.

\subsection{Results}

\begin{figure}
\includegraphics[width=246.0pt]{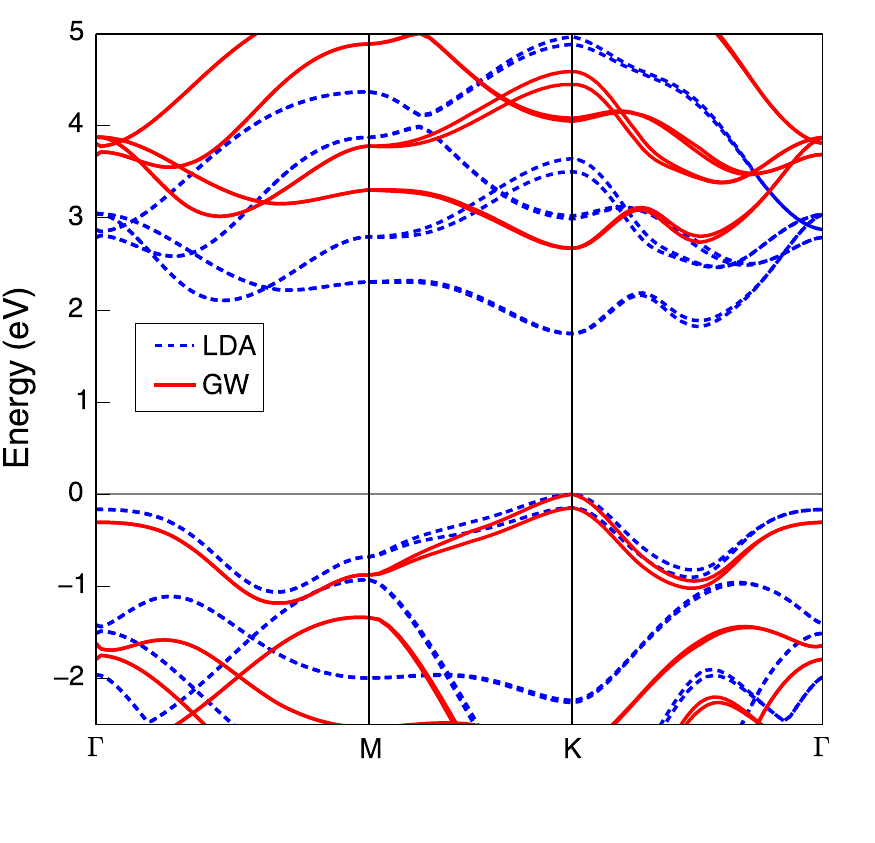}
\caption{\label{fig:bandstructure} (Color online) LDA (dashed blue curve) and $G_1W_0$ (solid red curve) band structure of monolayer \MoS2. }
\end{figure}

The bandstructure of monolayer \MoS2 at the LDA and $G_1W_0$ levels are shown in Fig.~\ref{fig:bandstructure}. We find that monolayer \MoS2 is a direct bandgap material at all levels of theory. The direct gap at the $K$ point increases from $1.71$~eV at the LDA level to $2.59$~eV at the $G_0W_0$ level to $2.67$~eV at the $G_1W_0$ level. The spin-orbit splitting of the valence band at $K$ is $0.15$~eV.

The GW correction varies by k-point. The largest correction to the gap is $1.2$~eV at the $M$ point, and the smallest is $0.96$~eV at the $K$ point. The GW correction also changes the effective masses, making the electron mass smaller than the hole mass. At the LDA level, the electron and hole effective masses at the $K$ point are $0.5m_0$ and $0.6m_0$ respectively. At the $G_1W_0$ level, the electron and hole effective masses are $0.4m_0$ and $0.2m_0$ respectively.

\subsubsection{\label{sect:qp_comparison}Comparison with other Calculations}

There is significant disagreement on the electronic structure of monolayer \MoS2,
including whether it has a direct or indirect gap, at various levels of theory,
though it is well-known that the experimental gap is direct~\cite{mak10}.
We compare our results with previous GW calculations on monolayer MoS2 in
Table~\ref{tab:gw}. Several calculations~\cite{huser13, shi13} find an indirect gap
from $\Gamma$ to $K$ the $G_0W_0$ level, and Shi \textit{et al.}~\cite{shi13} argue that self-consistently
updating G makes the gap direct. We find a direct gap at the $K$ point at all levels
of theory regardless of k-point sampling and the truncation of the Coulomb interaction.
Different k-points converge with respect to $N_b$ and $E_S$ at different rates, and the
$\Gamma$ point converges much more quickly than the $K$ point, so the indirect gap seen
in some calculations is likely an artifact of a too small dielectric cutoff.
Because the self-energy correction is larger at $\Gamma$ than at $K$, self-consistently
updating $G$ may fortuitously restore the direct gap in those calculations.



Besides convergence, the largest source of differences across previous GW calculations
 on monolayer MoS$_2$ is the use of a truncated Coulomb interaction. As discussed in
 section~\ref{sect:truncation} and also seen in Refs.~\cite{komsa12,huser13}, not using Coulomb truncation in a calculation with periodic boundary
 conditions results in over screening and decreases the QP gap by $100-300$~meV
 depending on the supercell size used.

\input{tab_gw_ff}

\section{Optical Properties}

\subsection{Computational Details and Convergence}

\subsubsection{\label{sect:opt_conv}False Convergence and Shift of the Electron-hole Continuum}

\begin{figure}
\includegraphics[width=246.0pt]{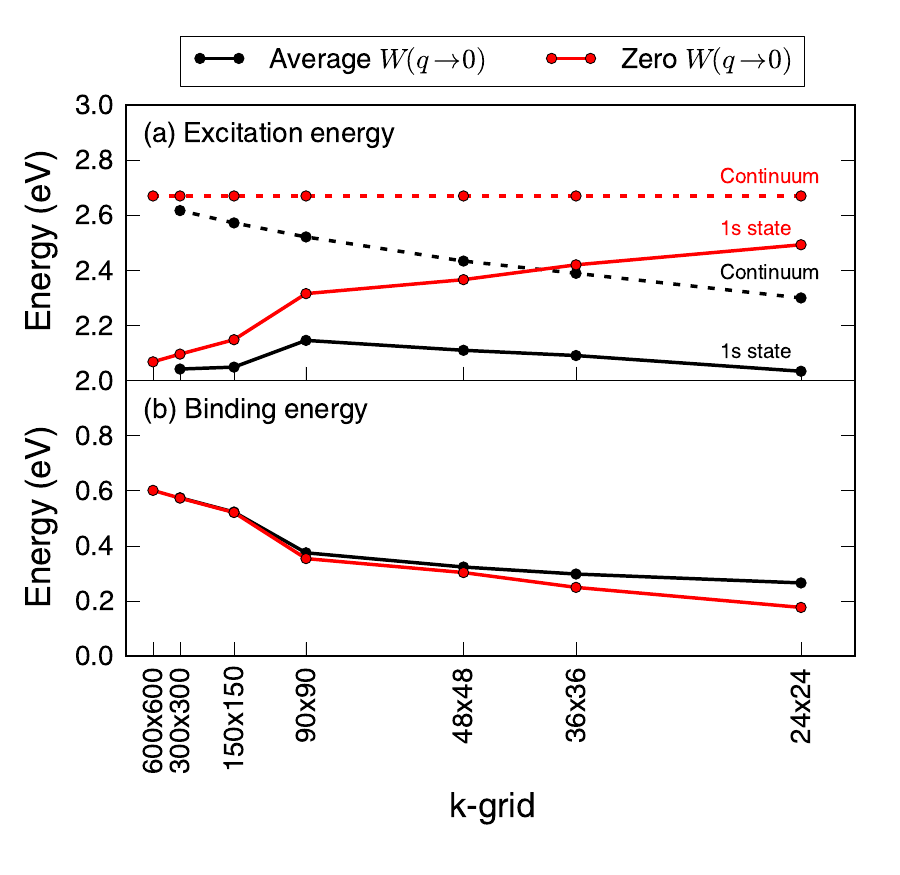}
\caption{\label{fig:q0} (Color online) (a) Convergence with respect to k-point
  sampling of the exciton continuum (dashed lines) and the 1st (1s) excitation
  energy (solid lines) for the A series of excitons when setting $W(q\rightarrow
  0)=0$ (red) or using $W^{\mathrm{avg}}(q\rightarrow 0)$ (black). (b)
  Convergence of the binding energy, defined as the difference between the
  continuum onset and the excitation energy of the 1st exciton in the A series
  when setting $W(q\rightarrow 0)=0$ (red) or using
  $W^{\mathrm{avg}}(q\rightarrow 0)$ (black). }
\end{figure}

As several works have noted, the optical properties of monolayer \MoS2,
 as calculated using the Bethe-Salpeter Equation (BSE) formalism,
 converge very slowly with respect to k-point sampling~\cite{komsa12,huser13,qiu13}.
 In reduced-dimensional systems, the screening varies rapidly as $\mathbf{q}$ approaches the long
 wavelength limit (See section II). Excitons at the $K$ point in \MoS2 are highly
 localized in momentum space, which means they are extended in real space, so most of the screening comes from the rapidly varying
 portion of $\epsilon_\mathrm{2D}(q)$. Hence, convergence with respect to k-point
 sampling is slow because it is necessary to resolve the fast changes in spatial dependence in screening.
 The extent of the exciton wavefunction in k-space is discussed in greater detail in
 Section IV.B.2. We find that a 300x300x1 k-grid is required to converge the exciton
 binding energy to within $0.1$~eV (Fig.~\ref{fig:q0}) for the lowest energy state. It is even more demanding for the excited exciton states.

The convergence of the excitation energies with k-point sampling varies depending
 on the treatment of the divergent term $W(\mathbf{q}{=}0)$. For semiconductors,
 the screened Coulomb interaction $W(\mathbf{q})$ diverges at $\mathbf{q}=0$, and it is common to avoid this
 divergence by replacing the screened interaction, $W(\mathbf{q}{=}0)$, with an average
 over a small region of the Brillouin zone~\cite{jdeslip11BGW,huser13} near $q=0$. We compare
 two different methods of treating the $\mathbf{q}=0$ term. In the first, we average the
 screened Coulomb interaction over a small volume in reciprocal space around
 $\mathbf{q}=0$. That is, we replace the divergent term, $W_{00}(\mathbf{q}{\rightarrow}0)$, with
\begin{equation}
W_{00}^{\mathrm{avg}}(\mathbf{q})=\frac{N_\mathbf{q}V}{2\pi}\int_\mathrm{cell} \mathrm{d}^2q\; W_{00}(\mathbf{q}),
\end{equation}
where ``cell'' indicates an integral over the volume of the Voronoi cell around
$\mathbf{q}=0$, $N_\mathbf{q}$ is the total number of q-points, $V$ is the
volume of the unit cell in real-space, and $W_{00}$ refers to the divergent
``head'' element, $\mathbf{G}{=}\mathbf{G}'{=}0$.

This averaging treatment results in faster convergence of the excitation energies with k-point
 sampling, but the convergence is non-variational -- i.e. the excitation
 energy initially increases with k-point sampling (Fig.~\ref{fig:q0}~(a)).
 The non-variational convergence occurs because replacing $W(\mathbf{q}{=}0)$
 with its average means that a k-point-dependent value is being added to
 the diagonal of the BSE matrix, which is equivalent to shifting the
 exciton continuum by $W^{\mathrm{avg}}(\mathbf{q}{=}0)$.

We emphasize that, while the widely-used averaging scheme is useful for
 improving the convergence of the excitation energies, it may lead
 to misleading binding energies, defined as the difference between the
 optical gap and the continuum of optical transitions. From Fig.~\ref{fig:q0}, the
 excitation energy from a relatively coarse 24x24x1 k-grid appears to
 agree better with experiment than finer k-grids, but if the shift
 to the continuum energy is taken into account, the binding energy
 is only $0.2$~eV. As k-grid sampling increases, the continuum energy
 increases linearly with $1/\sqrt{N_\mathbf{k}}$. Even more surprisingly,
 the excitation energy varies in a non-uniform way, and increases until we hit
 a k-grid finer than about $90$x$90$. For k-grids finer than this, we start
 to sample $\mathbf{q}$ vectors before the peak in the quasi-2D dielectric
 screening. Because the excitons are fairly spread out in real space, it is
 necessary to sample very small wave vectors to capture the small screenings
 associated with these lengh scales.

In an alternative treatment of $\mathbf{q}=0$, we fix the exciton continuum at the QP gap
 ($E_c-E_v$) by setting $W_{00}(\mathbf{q}{=}0)=0$, which is the value of
 $W^{\mathrm{avg}}$ in the limit of infinite k-points. In this scheme,
 the excitation energies converge slower with respect to k-point sampling,
 but the continuum does not move and the convergence is variational.
 There is again a kink in the convergence of the excitation energy around
 $90$x$90$, which comes from increased sampling in the small $\mathbf{q}$ region.
 If we define the binding energy as the difference between the excitation energy
 and the onset of the electron-hole or exciton continuum, the binding energy converges at roughly
 the same rate regardless of the treatment of $W(\mathbf{q}{=}0)$.

Therefore, even though the comonly-used averaging scheme of the screened Coulomb
 interaction typically converges the optical excitation faster, it does so by
 moving the continuum of optical excitations and introduces errors in both the
 excitonic wave functions and the energies of higher excited exciton states. This is
 particularly important if one is interested in properties such as the radius of
 the excitonic wave function OR the energies and characters of excited excitonic states.

As a final remark, we note that the fact that the exciton is tightly localized in k-space
 reduces the dielectric cutoff, $E_S$, needed to capture the screening for exciton calculations as opposed to those for those for QP energies. As seen
 in Fig.~\ref{fig:eps2D}, for $q < \angs{0.1}^{-1}$, the screening is the same for
 $E_S=8$~Ry and $E_S=35$~Ry. Indeed, when we reduce the cutoff from $35$~Ry to
 $8$~Ry the binding energies of the first $40$ excitonic states change by less
 than $10$~meV.

\subsubsection{Computational Details}

In this section, we describe the techniques that allows us to solve the BSE with a very dense k-point sampling and include spin-orbit effects.

In Fig.~\ref{fig:q0}, we explicitly solve the BSE on k-grids with up to
600x600x1 k-points in the full Brillouin zone. However, to save computational
cost, we only included k-points within $\angs{0.2}^{-1}$ of the $K$ point. This
is reasonable for testing convergence, since more than 99\% of states contributing to the lowest energy exciton fall within $\angs{0.1}^{-1}$ of the K point. To obtain the entire optical spectrum, however, it is necessary to consider the entire Brillouin zone using a k-point sampling of at least 300x300x1, which is very computationally demanding.

Rohlfing and Louie~\cite{rohlfing98} originally proposed an interpolation scheme to eliminate this computational bottleneck using two distinct k-grids, a coarse one where the matrix elements for the BSE are calculated, and a fine one onto which the matrix elements are interpolated and on which the BSE Hamiltonian is diagonalized. However, this interpolation scheme is no longer accurate for quasi-2D materials since the dielectric matrix has a lot of structure for small $\mathbf{q}$'s, contrary to the case for bulk systems.

Here, we modify this interpolation scheme to fully capture these fast variations in $\epsilon^{-1}_{00}(\mathbf{q})$ for small $\mathbf{q}$'s. As in the original scheme, we use two k-grids: a fine 300x300x1 k-grid and a coarse 24x24x1 k-grid where we explicitly calculate the BSE matrix elements between all coarse k-points $\mathbf{k}_\mathrm{co}$. However, in addition to these matrix elements, we also calculate transitions from each coarse k-point to a number of fine k-points that form a cluster around each coarse k-point. We call this second set of matrix elements that capture small $\mathbf{q}$'s the \textit{cluster matrix elements}.

When we perform the interpolation of the matrix elements from the coarse to the
fine k-grid, we use the original scheme from Rohlfing and
Louie~\cite{rohlfing98} if a particular transition has a wave vector
$\mathbf{q}=\mathbf{k}_\mathrm{fi}-\mathbf{k}'_\mathrm{fi}$ larger than a given
threshold. Otherwise, we use the cluster matrix element. This interpolation
scheme explicitly captures the fast variation in screening at small
$\mathbf{q}$'s, and the resulting excitation energies of the first 20 exciton
states are within $20$~meV of excitation energies found by explicitly calculating the BSE matrix on a 300x300x1 k-grid.

In addition to k-point sampling, it is also important to consider spin-orbit interactions in the optical absorption spectrum. If one directly solves the BSE on a relativistic basis set that includes spin-orbit interactions, the time to diagonalize the BSE grows by a factor of $64$ compared to the non-relativistic case and would not allow one to use such fine k-point sampling. An alternative scheme to include spin-orbit interactions is therefore desirable.

Our solution is to take advantage of the facts that (1) spin-orbit
 splitting is smaller than the exciton binding energy; and (2) spin along the z-axis is a good
 quantum number at the $K$ and $K'$ points for monolayer TMDs~\cite{xiao12}. This allows us to efficiently
 include spin-orbit effects as a perturbation. We perform both a spin-unpolarized
 DFT calculation, which is used as the starting wave functions for our GW calculation, and a
 non-collinear calculation with spin-orbit interactions included. We approximate the first-order
 spin-orbit correction to the GW quasiparticle energies to be the difference between the two
 Kohn-Sham eigenvalues. That is, we take
 $\Delta \varepsilon_{\mathrm{GW}}^{\mathrm{SO}}(n\mathbf{k}\sigma) \approx \Delta \varepsilon_{\mathrm{LDA}}^{\mathrm{SO}}(n\mathbf{k}\sigma) \equiv \varepsilon^{\mathrm{non-col}}(n\mathbf{k}\sigma) - \varepsilon^{\mathrm{unpol}}(n\mathbf{k})$,
 where $\sigma$ is the spinor index of the states in the non-collinear calculation.
 This is a reasonable approximation since the overlaps between the spinor wave functions and
 the scalar wave functions are exactly $1$ at $K$ and greater than $0.7$ in other regions
 with spin-orbit splitting, in our LDA calculation.

To obtain the absorbance with spin-orbit interaction, we apply a first-order perturbation
 theory to the solution of the Bethe-Salpeter equation, which is justifiable because the
 quasiparticle gap ($\sim 2.7$~eV) is much larger than the spin-orbit splitting ($\sim 150$~meV).
 Each excitonic state $\ket{S}$ can be expanded as a linear combination of pairs of
 single-particle valence and conduction band states as
\begin{equation}
\ket{S} = \sum_{vc\mathbf{k}}A^{S}_{vc\mathbf{k}}\ket{vc\mathbf{k}}.
\end{equation}

We want to calculate the spin-orbit corrected exciton energies
 $\Omega^{S}_{\sigma} = \Omega^{S} + \Delta\Omega^{S}_{\sigma}$, where $\Omega^{S}$ is the
 energy of the $\ket{S}$-state, neglecting spin-orbit, and $\Delta\Omega^{S}_{\sigma}$ is the
 first-order energy correction,
\begin{align}
\Delta\Omega^{S}_{\sigma} &\equiv \braket{ S | H^{\mathrm{SO}}_{\sigma} | S } \\
&= \sum_{vc\mathbf{k}}\sum_{v'c'\mathbf{k}'}
    \left( A^{S}_{v'c'\mathbf{k}'} \right )^* A^{S}_{vc\mathbf{k}}
    \braket{ v'c'\mathbf{k}' | H^{\mathrm{SO}}_{\sigma} | vc\mathbf{k} }, \notag
\end{align}
 where the spin-orbit Hamiltonian, $H^{\mathrm{SO}}$, is block-diagonal in the spin-index
 $\sigma$ and $H^{\mathrm{SO}}_{\sigma}$ is a block of the spin-orbit Hamiltonian for the spin $\sigma$.

We assume that $H^{\mathrm{SO}}_{\sigma}$ is diagonal in the $\ket{vc\mathbf{k}}$ basis,
 which is valid due to the large overlap between the spinor and scalar wave functions.
 Then, the spin-orbit correction to the excited-state energies becomes
\begin{equation}
\Delta\Omega^{S}_{\sigma} = \sum_{vc\mathbf{k}}\vert A^{S}_{vc\mathbf{k}}\vert^{2}\Delta \varepsilon^{\mathrm{SO}}_{vc\mathbf{k}\sigma}
\end{equation}
where $\Delta \varepsilon^{\mathrm{SO}}_{vc\mathbf{k}\sigma}$ are the spin-orbit corrected
 differences in energy between the valence and conduction states
\begin{align}
 \Delta \varepsilon^{\mathrm{SO}}_{vc\mathbf{k}\sigma} &=
 (\varepsilon_{\mathrm{GW}}(c\mathbf{k}) +  \Delta\varepsilon^{\mathrm{SO}}_{\mathrm{GW}}(c\mathbf{k}\sigma)) \notag \\
 &- (\varepsilon_{\mathrm{GW}}(v\mathbf{k}) + \Delta\varepsilon^{\mathrm{SO}}_{\mathrm{GW}}(v\mathbf{k}\sigma)) .
\end{align}

Finally, the imaginary part of the dielectric function with spin-orbit interactions is
 calculated using the spin-orbit corrected exciton energies,
\begin{equation}
\varepsilon_{2}(\omega)=\frac{16\pi^{2}e^{2}}{\omega^{2}}\sum_{S\sigma}\vert\mathbf{e}\cdot\bra{0}\mathbf{v}\ket{S\sigma}\vert^{2}\delta(\omega-\Omega^{S}_\sigma) ,
\end{equation}
 where $\mathbf{e}$ is the polarization of the incoming light, $\mathbf{v}$ is the
 velocity operator, and $\ket{S\sigma}=\ket{S}$.

\subsection{Optical Spectrum}

\begin{figure}
    \includegraphics[width=246.0pt]{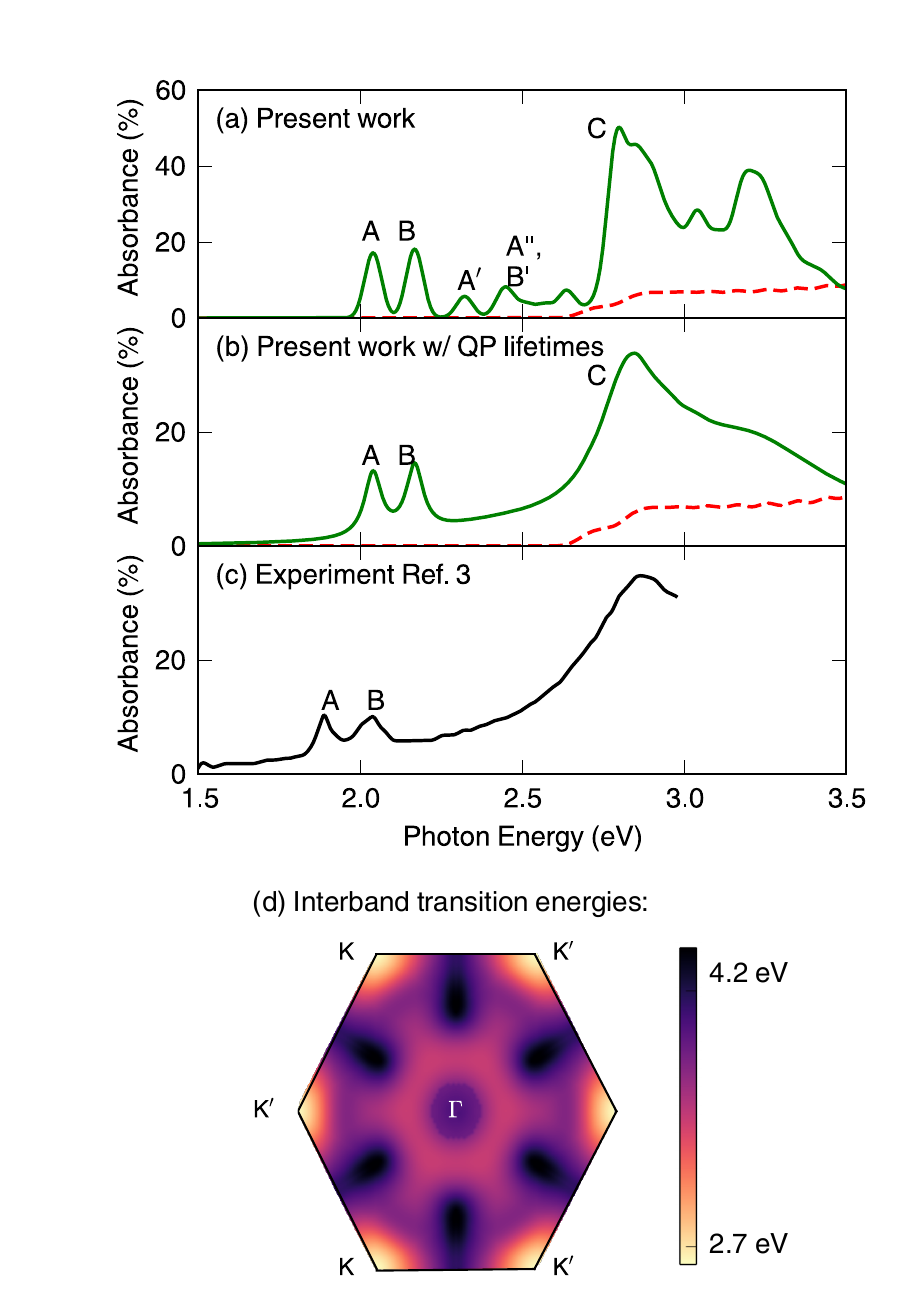}
    \caption{\label{fig:absp} (Color online) (a) Absorption spectra of \MoS2 without (dashed red curve) and with (solid green curve) electron-hole interactions using a constant broadening of $25$~meV. (b) Same calculated data as in Fig.~\ref{fig:absp}~(a), but using an \textit{ab initio} broadening based on the electron-phonon interactions\cite{marini08,li13}. (c) Experimental absorbance\cite{mak10}. (d) Direct valence to conduction band transition energies in the 1st Brillouin zone.}
\end{figure}

The absorption spectrum of monolayer \MoS2 with and without electron-hole interactions is shown in Fig.~\ref{fig:absp}. The lowest energy exciton, which forms peak A in the spectrum, has a binding energy of $0.63$~eV. Peaks A and B are spin-orbit split states that arise from excitons forming from transitions between the spin-orbit split valence band maximum and the conduction band minimum at the $K$ and $K'$ points in the Brillouin zone. Both A and B have bright excited states, which we label A', B', etc. The peak A'' overlaps with peak B'. We also see a large peak, which we label peak C, near the continuum onset at $2.7$~eV.

The lowest interband transition energies, i.e. the energies of direct transitions from the valence band to the conduction band throughout the Brillouin zone, are shown in Fig.~\ref{fig:absp}(d). The deepest valleys are parabolic valleys at $K$ and $K'$ points, which give rise to the A and B series of excitons. There is also a shallower Mexican-hat shaped valley around the $\Gamma$ point. Transitions from this Mexican hat valley give rise to peak C and its excited states.

The fine features due to excited states of peaks A and B, which appear in our calculated spectra, are broadened out in the experimental spectra. This is a signature of lifetime effects due to electron-phonon and other interactions. We account for the electron-phonon lifetime effects in our calculation following Marini~\cite{marini08}, and the result is plotted in Fig.~\ref{fig:absp}~(b). We consider both emission and absorption of phonons at $T=300$~K, and we extrapolate the scattering rate for quasiparticle energies larger than those computed by Li \textit{et al.}~\cite{li13}. This leaves the A and B peaks relatively sharp, while broadening out the intermediate peaks between B and C, resulting in excellent agreement with experiment for peak shape and position and the magnitude of the absorbance.


\subsubsection{Comparison with other Calculations}

As with the QP bandgap, there is a wide range of disagreement in the literature about
 the binding energy of the exciton giving rise to peak A at the GW-BSE level, with values
 ranging an order of magnitude from $0.1-1.1$~eV. A comparison of values obtained in
 different calculations is given in Table~\ref{tab:bse}. There is, however, a smaller spread in the
 calculated values of the excitation energy of peak A. This is largely because errors
 which result in over screening or under screening tend to affect the QP gap and binding
 energy in opposite ways, resulting in a cancellation of error in the excitation energy.
 The main sources of difference across various BSE calculations in the literature are:
 (1) the k-grid sampling, as mentioned in Section~\ref{sect:opt_conv}; and
 (2) the truncation of the Coulomb interaction. Coulomb truncation is especially
 important to obtain the correct binding energy because, as
 seen in Fig.~\ref{fig:eps2D}~(a,b), Coulomb truncation mainly affects screening
 in the small-q region where the exciton wavefunction is sensitive because of its localization in k-space. For instance, Refs.~\cite{huser13}
 and~\cite{wirtz13} have both noted that very fine k-point sampling is required to converge
 the solution of the BSE, yet obtain drastically different results ($0.6$ and $0.15$~eV,
 respectively) for the binding energy.


\input{tab_bse}

\subsection{Excitonic Spectrum of Series A and Comparison with Rydberg Series}

\begin{figure}
 \includegraphics[width=246.0pt]{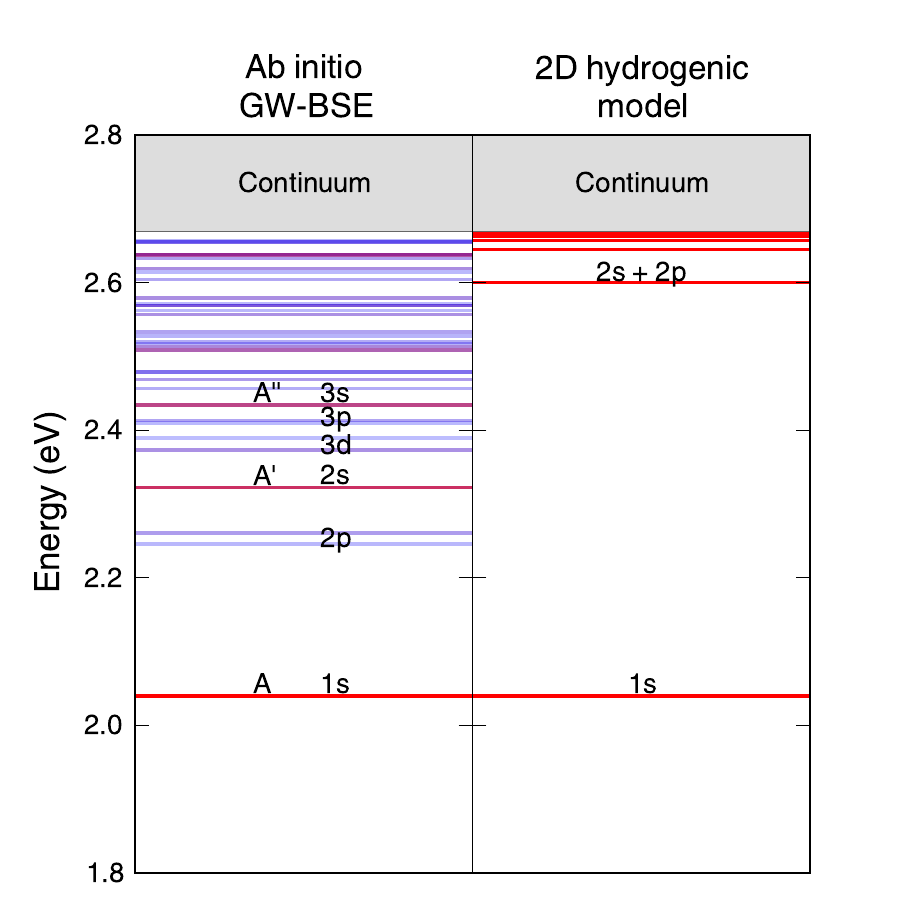}
\caption{\label{fig:xct_spec}Comparison of the exciton state energy levels for the A series obtained from \textit{ab initio} GW-BSE calculation (left) with an effective 2D hydrogenic model (right). Bright [dark] exciton states are represented by opaque red [translucent blue] lines.}
\end{figure}

\begin{figure}
    \includegraphics[width=246.0pt]{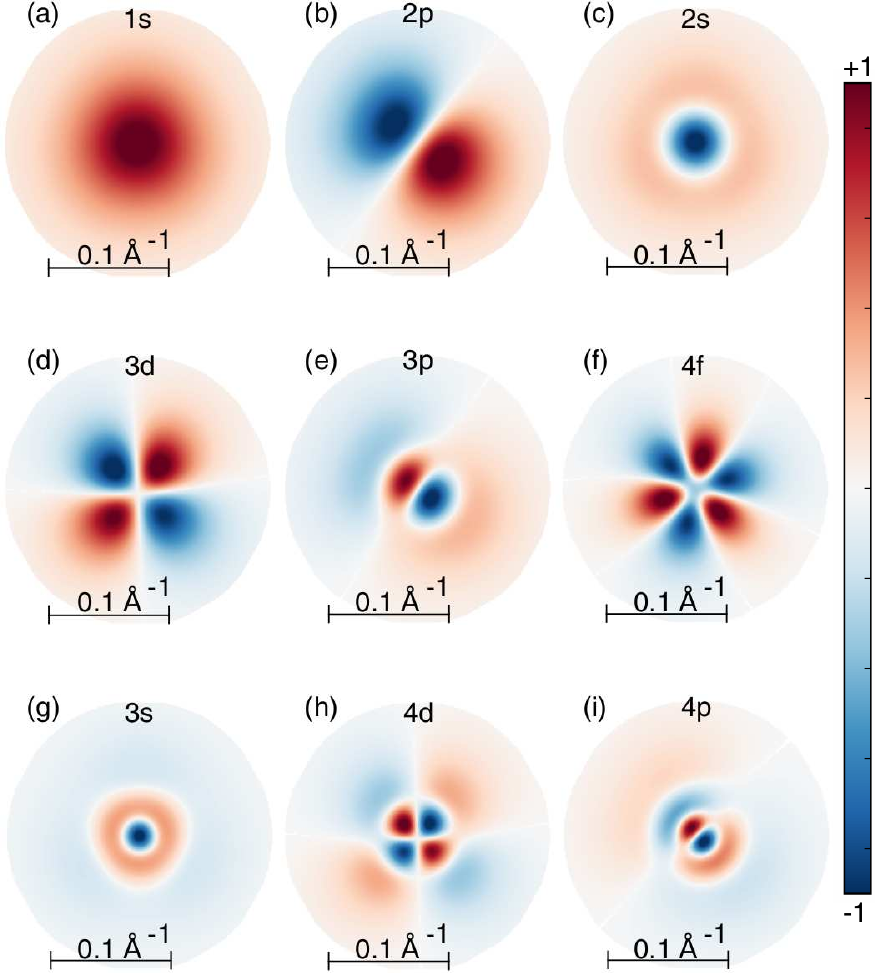}
    \caption{\label{fig:xct} (Color online) Electron-hole pair amplitudes of lowest energy exciton wavefunctions in reciprocal space for states (a) 1s, (b) 2p, (c) 2s, (d) 3d, (e) 3p, (f) 4f, (g) 3d, (h) 4d and (i) 4p. Each plot is centered around the $K$ point in the Brillouin zone.}
\end{figure}

We can obtain further insight of the structure of the excitonic states by comparing them
 to a 2D hydrogenic model. In Fig.~\ref{fig:xct_spec}, we plot the energies of the excitons in the
 series A obtained from our GW-BSE calculation with those from an effective 2D hydrogenic model
 $H_\mathrm{hydrog} = -\frac{\nabla^2}{2m^*} + \frac{e^2}{\epsilon^* r}$.
 This effective model is built by fitting the effective
 dielectric constant $\epsilon^*$ to reproduce the binding energy of
 peak A. Because there is very little coupling between the $K$ and $K'$ valleys~\cite{qiu15}, the A and B
 series of excitons are both doubly degenerate, and so we focus here on the states in the
 A series coming from a single valley.

As previously  noted\cite{qiu13,ye14,zhu14,chernikov14}, the hydrogenic model deviates from the \textit{ab initio} results in two significant ways:
 (1) first, the binding energies of excited states are much larger than expected from a 2D hydrogenic
 model; and (2) states with higher angular momentum have a larger binding energy than states with
 lower angular momentum. Additionally, there is also some splitting of states with the same
 angular momentum, such as $2p$ and $3d$, due to the trigonal warping of the \MoS2 bandstructure
 at the $K$ and $K'$ valleys. The $f$ states do not split because they have the same three-fold
 symmetry as the bandstructure. Although the excitation energies of the solutions of the BSE
 deviate from the hydrogenic model, for simplicity, we still label the states as $1s$, $2s$, $2p$, etc., using
 the same notation as a 2D hydrogenic model, based on the number of radial and azimuthal
 nodes in the envelope function of the exciton wavefunction.

To understand the physical reasons for these differences between the hydrogenic model and the \textit{ab initio} calculation, we will first analyze the character of the excitonic wave functions and see how the
 actual \textit{ab initio} and \textbf{q}-dependent screening differs from the hydrogenic model.
 Each excitonic state $\ket{S}$ can be expressed as a linear combination of the electron-hole
 transitions $|vc\mathbf{k}\rangle$,
\begin{equation}
 \ket{S} = \sum_{vc\mathbf{k}} A^{S}_{vc\mathbf{k}} \ket{vc\mathbf{k}} .
\end{equation}

The coefficients $A^{S}_{vc\mathbf{k}}$ describe the envelope function or electron-hole pair amplitude of the exciton wavefunction in
 reciprocal-space. The envelopes of the wavefunctions of the first few states in the $A$ series
 of excitons are plotted in Fig.~\ref{fig:xct}. The plots are centered around the $K$ point in the
 Brillouin zone. The nodal structure of the envelope function of the states is apparent from this plot. The plots also show
 that the excitonic wavefunctions are highly localized in k-space, with most transitions falling
 within $\angs{0.1}^{-1}$ or about $5\%$ of the Brillouin zone. In fact, this is well
 within the region of fast variation in screening seen in Fig.~\ref{fig:eps2D} and explains
 for the most part why convergence with respect to k-point sampling is so slow, since the k-point
 sampling must be fine enough to resolve \textit{both} the region before the peak in the dielectric
 screening and the nodal structure of the exciton wavefunctions.

\begin{figure}
    \includegraphics[width=246.0pt]{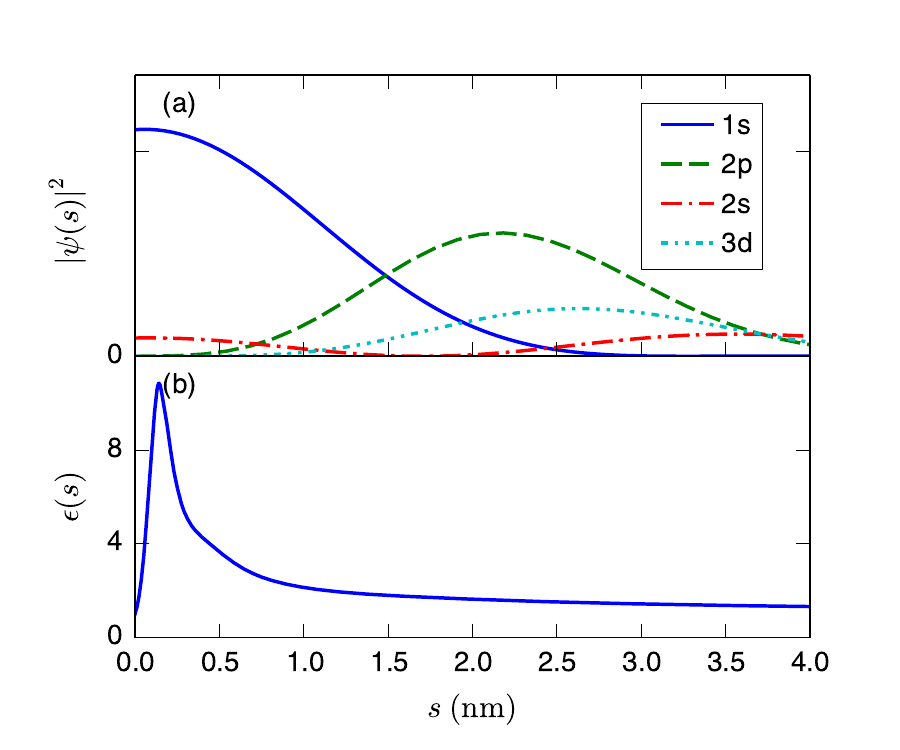}
    \caption{\label{fig:rxct} (Color online) (a) Modulus squared of the exciton wavefunction in real-space for the states 1s (solid blue line), 2s (red line with dash and dot), 2p (green dashed line), and 3d (cyan line with dash and two dots). (b) The effective 2D dielectric function over the same range in real space. }
\end{figure}

The deviations of the results of the \textit{ab initio}
calculation from those of the hydrogenic model may now be understood. If we compare the real-space screening $\epsilon(s)$
 with the envelope of the exciton wavefunctions in real space, as shown in
 Fig.~\ref{fig:rxct}, it is clear that the varying distribution of the
 wavefunction in real-space results in different states experiencing different
 screening and therefore different effective electron-hole interaction. In general, states with larger principal quantum number $n$ have a larger binding energy than expected from the hydrogenic model 
 because they have a larger radius and are thus less screened than states with smaller radii.
 Similarly, states with larger angular momentum quantum number are more strongly bound than in the model because
 there is a node in the wavefunction where screening is strongest.

Therefore, this effective state-dependent screening explains why
 (1) excited excitonic states, such as $2s$ and $3s$, appear lower in energy than
 what is predicted by a 2D hydrogenic model (which assumes a constant dielectric
 constant), and (2) degenerate states with the same principal quantum number $n$
 split, and the excitation energy for states with higher angular momentum is lower.



\section{Conclusion}

 In summary, we find that many-body effects, namely, the electron-electron and electron-hole interactions for quasiparticle and optical excitations, in MoS$_2$ are well-described by the GW-BSE method, which gives results in good agreement with experimental optical spectra and conclusions about the bandgap. We find that, for \MoS2, \GoWo/ results do not differ qualitatively from \scGWo/, as has been previously claimed. Instead, variations in GW-BSE results in the literature arise largely from different treatments of the long-range Coulomb interaction in periodic supercell calculations and convergence of k-grid sampling and cutoffs for the dielectric screening. We find that truncating the Coulomb interaction to prevent artificial over screening from periodic images is essential to obtain accurate results. The 2D nature of the system also gives rise to strong spatial variations in screening, which must be captured by very fine k-point sampling. The sharpest variation in screening is at small q-vectors (q$\lesssim\pi/d$, where $d$ is the layer thickness), where the screening rapidly vanishes as the wave vector $\mathbf{q}$ approaches zero. Even finer k-point sampling is required to converge the BSE, as the exciton electron-hole amplitude functions in \MoS2 are tightly localized in k-space. Finally, a large energy cutoff for the dielectric matrix is required to capture the spatial variation associated with the different characters of the VBM and CBM of \MoS2, and a correspondingly large number of empty states is required to avoid artificially truncating the dielectric matrix and capture the nearly continuous states arising from using a large vacuum. These are general conclusions that can be applied to GW-BSE calculations on any semiconductor in low dimensions.

We thank L. Yang and T. Cao for discussions. D. Y. Q. acknowledges support from the NSF Graduate Research Fellowship Grant No. DGE 1106400. Structural study and the work on calculating the electron-phonon interaction effects on the optical spectra were supported by NSF Grant No. DMR15-1508412. The GW-BSE calculations were supported by the Theory Program at the Lawrence Berkeley National Lab through the Office of Basic Energy Sciences, U.S. Department of Energy under Contract No. DE-AC02-05CH11231. This research used resources of the National Energy Research Scientific Computing Center, which is supported by the Office of Science of the U.S. Department of Energy, and the Extreme Science and Engineering Discovery Environment (XSEDE), which is supported by National Science Foundation grant number ACI-1053575.

\bibliography{MoS2}
\bibliographystyle{apsrev4-1}

\end{document}

%% file: tab_gw_ff.tex
\begin{table*}
\caption{\label{tab:gw}Comparison of smallest quasiparticle band gap
  ($E_\mathrm{gap,min}^{GW}$) and the QP gap at the K point (E$_{\mathrm{gap},K}^{GW}$) from a selection of different
  GW calculations on monolayer MoS$_2$. The calculations differ by the use of the truncated Coulomb
  interaction, the level of self-consistency, the method for including dynamical effects in the
  polarizability and the mean field starting point, including the DFT
  functional and the in-plane lattice constant ($a$), as well as convergence
  parameters. The compared convergence parameters are: use of Coulomb truncation, supercell size
  along the aperiodic direction ($L_z$), k-grid size, the
  energy cutoff for the dielectric matrix ($E_S$) and the number of bands
  included in the summation in the polarizability and the Coulomb-hole term in
  the self energy ($N_b$). The methods for describing dynamical effects in the polarizability
  (Freq. Dep.) are the Hybertsen-Louie Generalized Plasmon Pole (HL) model~\cite{hybertsen86},
  the Godby-Needs Plasmon Pole model (GN)~\cite{godby89}, or explicit calculation of the full
  frequency dielectric matrix (FF).}

\begin{ruledtabular}
\begin{tabular*}{\textwidth}{
@{\extracolsep{\fill}}lccccccccccc@{}}

~ & \multicolumn{6}{c}{Convergence Parameters} & \multicolumn{2}{c}{Starting Meanfield} &\multicolumn{3}{c}{QP Gaps} \\
\cline{2-7} \cline{8-9} \cline{10-12}  \\[-0.9em]
~ &  \begin{tabular}{@{}c@{}} Coulomb\\Trunc.\end{tabular} & \begin{tabular}{@{}c@{}} $L_z$ \\(\AA)\end{tabular} 
& k-grid & \begin{tabular}{@{}c@{}}$E_S$\\(Ry)\end{tabular} & $N_b$ & \begin{tabular}{@{}c@{}}Freq.\\Dep.\end{tabular}
& DFT & $a$ (\AA) & \begin{tabular}{@{}c@{}}E$_{\mathrm{gap},K}^{GW}$\\(eV)\end{tabular}
& \begin{tabular}{@{}c@{}}E$_\mathrm{gap,min}^{GW}$\\(eV)\end{tabular} & \begin{tabular}{@{}c@{}} Direct\\Gap\end{tabular}  \\

\cline{2-7} \cline{8-9} \cline{10-12} \\[-0.9em]
 \begin{tabular}{@{}c@{}}Present\\Work (G$_{1}$W$_{0}$)\end{tabular}
 &  \begin{tabular}{@{}c@{}}Y\\Y\end{tabular} & 
 \begin{tabular}{@{}c@{}}25\\25\end{tabular}
   &  \begin{tabular}{@{}c@{}}24x24x1\\24x24x1\end{tabular} & 
 \begin{tabular}{@{}c@{}}35\\35\end{tabular} &  \begin{tabular}{@{}c@{}}6000\\6000\end{tabular} &
  \begin{tabular}{@{}c@{}}HL\\FF\end{tabular} &  \begin{tabular}{@{}c@{}}LDA\\LDA\end{tabular} & 
 \begin{tabular}{@{}c@{}}3.15\\3.15\end{tabular} &  \begin{tabular}{@{}c@{}}2.67\\2.54\end{tabular}
     & 
 \begin{tabular}{@{}c@{}}2.67\\2.54\end{tabular} & \begin{tabular}{@{}c@{}}Y\\Y\end{tabular} \\ 
\\

G$_{1}$W$_{0}$\cite{qiu13} & Y & 25 & 24x24x1 & 35 & 6000 & HL & LDA & 3.15 & 2.7 & 2.7 & Y \\

G$_{1}$W$_{0}$\cite{qiu13} & Y & 25 & 12x12x1 & 35 & 6000 & HL & LDA & 3.15 & 2.84 & 2.84 & Y \\

\GoWo/\cite{soklaski14} & Y & 25 & 24x24x1 & 35 & 6000 & HL & PBE & 3.18 & 2.63 & 2.63 & Y \\

\GoWo/\cite{huser13} & Y & 23 & 45x45x1 & 3.7 & 200 & GN & LDA & 3.16 &2.77 & 2.58 &N\footnote{ Gap from $\Gamma\rightarrow K$} \\

\GoWo/\cite{wirtz13} & N & 24 & 18x18x1 & 2\footnote{$E_S$ estimated from supercell size and number of reported
  G-vectors in dielectric matrix (50).} & 200 & GN & LDA & 3.15 & 2.41 & 2.41 & Y  \\

\GoWo/\cite{komsa12} & N & 20 & 12x12x1 & 15 & 120\footnote{Number of bands estimated from supercell size and
  reported energy of highest band.}& FF & PBE & 3.18 & 2.60 & 2.60 & Y  \\

\GoWo/\cite{shi13} & N & 19 & 12x12x1 & 22 & 197 &  FF & PBE & 3.16 & 2.60 & 2.49 & N \\

\scGWo/\cite{shi13} & N & 19 & 12x12x1 & 22 & 197 & FF &PBE & 3.16 & 2.80 & 2.80 & Y \\

\GoWo/\cite{ashwin12} & N & 15 & 6x6x1 & 20 & 96 & FF & HSE & 3.18 & 2.82 & 2.82 & Y \\

\scGWo/\cite{conley13} & N & 9 & -- & 7 & -- & FF & PBE & 3.19 & 2.40 & 2.40 & Y  \\


QSGW\cite{lambrecht12} & N & 19 & 8x8x2 & -- & --& FF & LDA & -- & 2.76 & 2.76 & Y\\


\end{tabular*}
\end{ruledtabular}
\end{table*}

%% file: tab_bse.tex
\begin{table*}[htbp]
\caption{\label{tab:bse} Comparison of a selection of GW-BSE calculations for monolayer
  \MoS2, including the excitation energy ($\Omega$) of peaks A, B, A', B', and C, and
  the binding energy (E$_b$) of peak A, which is taken to be the difference between the QP gap and
  the excitation energy. If spin-orbit was not included in the
  calculation the excitation energy of peak B(B') is reported as the same as peak
  A(A'). Parameters affecting the calculation are k-grid sampling, the use of a
  truncated Coulomb interaction, and the number of valence ($N_v$) and
  conduction ($N_c$) states.
}

\begin{ruledtabular}
\begin{tabular*}{\textwidth}{
@{\extracolsep{\fill}}lcccccccccc@{}}

~ & \multicolumn{3}{c}{Convergence Parameters} & \multicolumn{2}{c}{Peak A} & Peak B & Peak A' &
  Peak B' & Peak C \\
\cline{2-4} \cline{5-6} \cline{7-7} \cline{8-8} \cline{9-9} \cline{10-10}\\[-0.9em]
~ & \begin{tabular}{@{}c@{}} Coulomb\\Trunc.\end{tabular} & k-grid & ($N_v$, $N_c$) 
& \begin{tabular}{@{}c@{}}$\Omega$\\(eV)\end{tabular} & \begin{tabular}{@{}c@{}}E$_b$\\(eV)\end{tabular} 
& \begin{tabular}{@{}c@{}}$\Omega$\\(eV)\end{tabular}
&\begin{tabular}{@{}c@{}}$\Omega$\\(eV)\end{tabular} & \begin{tabular}{@{}c@{}}$\Omega$\\(eV)\end{tabular}
& \begin{tabular}{@{}c@{}}$\Omega$\\(eV)\end{tabular} 
\\

\cline{2-4} \cline{5-6} \cline{7-7} \cline{8-8} \cline{9-9} \cline{10-10}\\[-0.9em]

Present Work & Y & 300x300 & (4, 4) & 2.04 & 0.63 & 2.17 & 2.32 & 2.45 & 2.7 \\ 
\\

Ref. \cite{qiu13} & Y & 300x300 & (4, 4) & 2.04 & 0.63 & 2.17 & 2.32 & 2.45 & 2.73 \\

Ref. \cite{qiu13} & Y & 72x72\footnote{k grid interpolated following Rohlfing and Louie.\cite{rohlfing00}} & (7, 8) & 1.88 & 0.96 & 2.02 & 2.20 & 2.32 & 2.54 \\

Ref. \cite{soklaski14} & Y & 60x60$^\mathrm{a}$ & (4, 4) & 1.94 & 0.62 & 2.08 & 2.4 & -- & 2.7 \\

Ref. \cite{huser13} & Y & 45x45 & (1, 1) & 2.2 & 0.6 & 2.2 & -- & -- & --\\

Ref. \cite{komsa12} & N\footnote{$E_b$ and QP gap
  are extrapolated to $L_z=\infty$} & 12x12 &  (--, --) & 1.9 & 1.1 & 1.9 & -- & -- &-- \\

Ref. \cite{wirtz13} & N & 51x51 & (--, --)& $\sim$2.2\;\; & 0.15 & $\sim$2.3\;\; & -- & -- & 3.0 \\

Ref. \cite{sanchez16} & N & 30x30 & (2, 4)& 2.0 & $\sim$0.7\;\; & 2.15 & -- & -- &  $\sim$2.95\;\; \\

Ref. \cite{palummo15} & N & 27x27 & (6, 6) & 2.03 & -- & 2.14 & -- & -- & $>$2.6 \\

Ref. \cite{bernardi13} & N & 16x16 & (6, 8) & 2.11 & -- & 2.25 & -- & -- & 2.55 \\

Ref. \cite{shi13} & N & 15x15 & (6, 8) & 2.22 & 0.54 & 2.22 & 2.5 & 2.5 & 3.0 \\

Ref. \cite{ashwin12} & N & 6x6 & (4, 8) & 1.78 & 1.04 & 1.96 & -- & -- & ~3.0 \\

Ref. \cite{conley13} & N & -- & (6, 8) & 1.86 & 0.56 & --& -- & -- & --\\

\end{tabular*}
\end{ruledtabular}
\end{table*}